\documentstyle[amssymb,aps,overcite,preprint]{revtex}

\begin{document}
\title{Probing the local structure: macromolecular combs in external fields }
\author{Raffaella Burioni (1), Davide Cassi (1) and Alexander Blumen (2)*}
\address{(1) Istituto Nazionale di Fisica della Materia-Unit\`{a} di Parma,\\
Dipartimento di Fisica, Universit\'{a} di Parma, Parco Area delle\\
Scienze 7/A, I-43100 Parma, Italy\\
(2) Theoretische Polymerphysik, Universit\"{a}t Freiburg, Hermann-Herder-\\
Stra\ss e 3, D-79104 Freiburg, Germany,\\
\thinspace * to whom proofs should be sent. Telephone-Number:\\
xx49-761-203-5905, Fax-Number: xx49-761-203-5906, e-mail:\\
blumen@physik.uni-freiburg.de }
\maketitle

\begin{abstract}
Recent experimental methods allow to monitor the response of macromolecules
to locally applied fields, complementing usual, mesoscopic techniques. Based
on the Rouse-model and its extension to generalized Gaussian structures
(GGS), we follow here the stretching of comb macromolecules under local
fields. This leads to a wealth of informations about the structure: Namely,
given the inhomogeneous architecture of combs, the dynamics and amount of
stretching depend strongly on the position of the monomer on which the
external fields act. We discuss both the theoretical and the experimental
implications of our findings, given that micromanipulations can be
supplemented by fluorescence measurements, which are very sensitive to
changes in the intramolecular distances.
\end{abstract}

\section{Introduction}

Due to numerous technological applications, understanding the structural and
dynamical properties of branched polymers and of polymer networks is of much
interest.\cite{T75, ME92} Here a very challenging question is to determine
how the topology affects the dynamical properties.\cite{T75, ME92, LSB99,
BKB00} Apart from the classical, bulk determination of the relaxation
moduli, nowadays one can also act microscopically on dilute polymers in
solution; for instance one can locally apply electrical fields on charged
polymers (polyelectrolytes, polyampholytes), magnetic fields on magnetizable
beads, or optical tweezers on dielectric spheres attached to the
macromolecules.\cite{QBC97, PSLC, W95, GK01} Now, the extension of the
macromolecule under such external fields depends, evidently, on the
underlying topology and, especially, on the site on which the field acts. In
previous works we have shown that this site dependence gets to be more
pronounced, when the structure is more ramified; thus dendrimers \cite
{BKB00, KBB01} show larger site differences than linear chains and regular
fractals\cite{Sch98, FSB99, SFB99, BJ02}.

In this paper we study the stretching of comb-polymers \cite{ME83, RT87,L91,
YMcLS94, F99} whose backbones are linear chains, and especially the
stretching of ring-backbones out of which sprout linear chains. As stressed,
we choose these systems in order to study in detail the differences
encountered when applying local fields to different constituents of rather
heterogeneous macromolecules; furthermore, we know that dynamical features
of comb-like structures can differ vastly, in the asymptotic limit, from the
behavior encountered in objects with regular topologies.\cite{CR92,BCV99,
BCV00}

As we proceed to show, modern measurements which allow to pinpoint the
external fields on local parts of large macromolecules lead, based on their
dynamical stretching, to a quite detailed picture of the underlying
connections inside the structure. Such changes in the distances inside the
macromolecule can be monitored directly, say by attaching a donor and an
acceptor chromophore to different parts of the structure and by following
the corresponding excitation transfer;\cite{KB84, RB89, SMB98, PSK98, BKK97}
due to the high sensitivity of the transfer on the mutual distance, the
method is very accurate, being of much use in polymer sciences. \cite{PSK98,
W99, RW01, FSW01, BBSM01, BBM, Y00, LP00, BKK97, SMB98}

In order to discuss the general situation and to stress the main features we
restrict ourselves to the basic ideas of the Rouse-model\cite{R53, DE86,
GK94}, as implemented in the generalized Gaussian-structures (GGS) scheme%
\cite{SB95, Sch98 }. Now using the GGS simplifies much the calculations,
given that in this scheme the effects of the hydrodynamic interactions and
of possible memory effects are neglected. Also neglected are local
geometrical aspects, such as the excluded volume and the stiffness of the
chains. Given, however, that our intention here is to show how much
additional information can be obtained from micromanipulations (as
contrasted to the usual bulk experiments) we prefer to work in the GGS
picture, while being well-aware of the fact that a quantitative comparison
to experiments must take additional features into account.

We recall that GGS consist of beads connected to each other by springs,
beads which feel the influence of the embedding medium through its
viscosity. The main theoretical advantage of centering on GGS is that they
allow to show the close interconnection between topology and dynamics.
Namely, as stated before in the GGS scheme many macroscopical observables of
polymer physics and of physical chemistry are simply related to the {\em %
eigenvalues}\cite{BKB00, DE86, T75, ME92, Sch98}{\em \ }of the connectivity
matrices of the GGS. However, these observables do not depend on the
corresponding eigenvectors. The $local$ probing of such structures, on the
other hand involves directly the {\em eigenvectors},\cite{BKB00, Sch98,
KBB01} fact which leads to a much wider class of possible dynamical
behaviors. In the case of comb-molecules, as we will show, letting the force
act closely to the tips or closely to the backbone of the structure leads to
markedly different responses. Possible ways to monitor such responses are
the tracking of the motion through fluorescent probes, as well as
nuclear-resonance and electronic energy transfer, the later methods being
very sensitive to changes in the relative distances inside the macromolecule.%
\cite{KB84, RB89, SMB98, PSK98, W99, RW01, FSW01, BBSM01,BBM, Y00, LP00,
BKK97}

The paper is structured as follows: In the next Section we discuss the
GGS-model and its implications for the dynamics. We recall in particular how
several basic, experimentally readily accessible quantities are related to
the properties of the connectivity matrix; in this Section we also show how
the positions on which the local forces act influence the stretching of the
macromolecules involved. Section III\ is devoted to the calculation of the
stretching of comb-molecules in external fields; in this Section we also
discuss our numerical findings related to the position-dependent dynamics
and give approximate analytical expressions for the response functions to
external fields. We focus in particular on the time-regime intermediate
between very small times (at which only small parts of the molecule are in
motion) and very long times (where the whole molecule moves); such
intermediate times are most revealing for the comb-like structures under
investigation. We conclude our work in Sec. IV with a discussion of results
and with indications about their implications for further theoretical and
experimental studies.

\section{Generalized Gaussian Structures}

In this section we consider the dynamics of Gaussian generalized structures
(GGS), whose simplest representation is the Rouse chain.\cite{R53} We follow
the usual development of the theory\cite{EM78, ME80, KMF90, SB95, DE86,
GK94, Sch98, FSB99, SFB99, GP92, AG91, F97}, while paying particular
attention to the extension of the GGS in external fields. As indicated
above, recent optical and mechanical developments allow one to
micromanipulate such GGS in solution.

We model the GGS as a complex consisting of $N$ beads, which are connected
to each other by harmonic springs. As usual, we assume that each monomer
experiences the friction coefficient $\zeta $ and that all beads move under
the influence of random forces, included here via the velocities ${\bf w}(t)$%
. Taking as usual the distribution of the ${\bf w}(t)$ to be Gaussian and
zero-centered, one obtains the following linearized Langevin equation for
the dynamics of the beads \cite{DE86, SB95, KMF90, GK94, GP92}, where one
denotes the coordinate of the $i$-th bead by ${\bf R}_{i}$ and the external
force acting on it by ${\bf F}_{i}$: 
\begin{equation}
\frac{\partial {\bf R}_{i}(t)}{\partial t}+\sigma \sum\limits_{j=1}^{N}A_{ij}%
{\bf R}_{j}(t)={\bf w}_{i}(t)+\frac{{\bf F}_{i}(t)}{\zeta }.  \label{A}
\end{equation}
Note that ${\bf A}=(A_{ij})$ in Eq. (\ref{A}) is the connectivity matrix\cite
{AG91, KMF90, SB95}. The matrix {\bf A }can be constructed by initially
setting all elements to zero and accounting for each bond between the
monomers $i$ and $j$ by increasing the diagonal elements $A_{ii}$ and $%
A_{jj} $ by +1 and the non-diagonal elements $A_{ij}$ and $A_{ji}$ by -1.
Note that in this way {\bf A} is a symmetric constant matrix. Furthermore $%
det\left( {\bf A}\right) =0$, as is evident by construction, and which
implies that (at least) one eigenvalue vanishes. More compactly, Eq. (\ref{A}%
) reads: 
\begin{equation}
\frac{\partial {\bf R}(t)}{\partial t}+\sigma {\bf AR}(t)={\bf w}(t)+\frac{%
{\bf F}(t)}{\zeta }  \label{B}
\end{equation}
with ${\bf R\equiv }({\bf R}_{1},{\bf R}_{2}\cdots ,{\bf R}_{n})^{T},$ ${\bf %
w\equiv }({\bf w}_{1},{\bf w}_{2}\cdots ,{\bf w}_{N})^{T}$ and{\bf \ }${\bf %
F\equiv }({\bf F}_{1},{\bf F}_{2},\cdots ,{\bf F}_{N})^{T}$, where $T$
denotes the transposed vector.

The solution of Eq. (\ref{B}) can be written as\cite{BKB00, KBB01} 
\begin{equation}
{\bf R}(t)=\int_{-\infty }^{t}dt^{\prime }\exp \left[ -\sigma (t-t^{\prime })%
{\bf A}\right] \left[ {\bf w}(t^{\prime })+\frac{{\bf F}(t^{\prime })}{\zeta 
}\right] .  \label{C}
\end{equation}
One can now diagonalize the connectivity matrix and has formally 
\begin{equation}
{\bf A=Q\Lambda Q}^{-1},  \label{D}
\end{equation}
where ${\bf \Lambda }$ is a diagonal matrix built from the eigenvalues of $%
{\bf A}$ and ${\bf Q}$\ is the matrix of the corresponding eigenvectors.
Inserting Eq. (\ref{D}) into Eq. (\ref{C}) leads to the mean displacement: 
\begin{equation}
\left\langle {\bf R}(t)\right\rangle =\int_{-\infty }^{t}dt^{\prime }{\bf Q}%
\exp \left[ -\sigma (t-t^{\prime }){\bf \Lambda }\right] {\bf Q}^{-1}\frac{%
{\bf F}(t^{\prime })}{\zeta }.  \label{E}
\end{equation}
In Eq. (\ref{E}) the average goes over the random velocities ${\bf w}$. We
now focus on the special case in which a constant external force is switched
on at $t=0$ and acts on the $m$-th bead only, i.e. ${\bf F}_{i}(t)={\bf F}%
_{0}\delta _{i,m}\sigma (t)$. The evaluation of $\left\langle
Y_{m}(t)\right\rangle $, the mean displacement of the $m$-th bead between
times 0 and $t$ is most readily performed by choosing the $y$-coordinate in
the direction of the force. We obtain then from Eq. (\ref{E}): 
\begin{eqnarray}
\left\langle Y_{m}(t)\right\rangle &=&\frac{F_{0}}{\zeta }%
\sum\limits_{i=1}^{N}\int_{0}^{t}dt^{\prime }Q_{mi}\exp \left[ -\sigma
\lambda _{i}(t-t^{\prime })\right] Q_{im}^{-1}  \label{Fa} \\
&=&\frac{F_{0}t}{N\zeta }+\frac{F_{0}}{\sigma \zeta }\sum%
\limits_{i=2}^{N}Q_{mi}\frac{1-\exp (-\sigma \lambda _{i}t)}{\lambda _{i}}%
Q_{im}^{-1}\text{,}  \nonumber
\end{eqnarray}
where we set (${\bf Q}^{-1})_{im}\equiv Q_{im}^{-1}$, noticed that these
quantities are independent of $t$ and hence performed the integration over $%
t $ in straightforward manner. Note that on the rhs of Eq. (\ref{Fa}) the
motion of the center of mass (CM) has separated automatically from the rest.%
\cite{BKB00, Sch98} The CM is characterized by the vanishing eigenvalue $%
\lambda _{1}=0$.

Eq. (\ref{Fa}) is the fundamental quantity for our further studies; one may
note that it depends both on the eigenvalues and on the eigenvectors of the
connectivity matrix {\bf A}. Before proceeding to analyse Eq. (\ref{Fa}) we
recall that averaging it also over all positions $m$ leads to an extremely
simple form, namely to\cite{Sch98} 
\begin{equation}
\left\langle \left\langle Y(t)\right\rangle \right\rangle =\frac{F_{0}t}{%
N\zeta }+\frac{F_{0}}{\sigma N\zeta }\sum\limits_{i=2}^{N}\frac{1-\exp
(-\sigma \lambda _{i}t)}{\lambda _{i}}.  \label{G}
\end{equation}
In Eq. (\ref{G}) one should remark that only the eigenvalues of the matrix 
{\bf A}, but not the eigenvectors are involved. Eq. (\ref{G}) is, in fact,
related to the dynamical response function (relaxation modulus) $G(t)$ of
the structure 
\begin{equation}
G(t)=\frac{1}{N}\sum\limits_{i=1}^{N}e^{-\sigma \lambda _{i}t}  \label{H}
\end{equation}
through: 
\begin{equation}
\left\langle \left\langle Y(t)\right\rangle \right\rangle =\frac{F_{0}}{%
\zeta }\int_{0}^{t}G(\widetilde{t})d\widetilde{t}.  \label{I}
\end{equation}
This relation shows that $G(t)$ plays the role of a fundamental dynamical
expression. In fact, $G(t)$ is also connected to other basic observables,
such as $G^{\prime }(\omega )$, the storage modulus and $G^{\prime \prime
}(\omega )$, the loss modulus\cite{KMF90, DE86, F97, GB01}. Now $G^{\prime
}(\omega )$ and $G^{\prime \prime }(\omega )$ are proportional to the real
and the imaginary part of the Fourier-transformed $i\omega G(2t);$ one has
namely 
\begin{equation}
G^{\prime }(\omega )=A\sum\limits_{i=2}^{N}\frac{\omega ^{2}}{\omega
^{2}+(2\sigma \lambda _{i})^{2}}  \label{J}
\end{equation}
for the storage modulus and 
\begin{equation}
G^{\prime \prime }(\omega )=A\sum\limits_{i=2}^{N}\frac{2\sigma \omega
\lambda _{i}}{\omega ^{2}+(2\sigma \lambda _{i})^{2}}  \label{K}
\end{equation}
for the loss modulus, where in Eqs. (\ref{J}) and (\ref{K}) for a given
sample and at a fixed temperature $T$ the prefactor $A$ is a constant.

In the next section we will discuss the stretching of comb-rings (a special
family of comb-like molecules) in the framework of the GGS dynamics
discussed here. The stretching $\delta Y(t)$ is given by subtracting from
Eq. (\ref{Fa}) the motion of the CM, $i.e.:$ 
\begin{eqnarray}
\left\langle \delta Y_{m}(t)\right\rangle =\left\langle
Y_{m}(t)\right\rangle -\frac{F_{0}t}{N\zeta } &&  \label{L} \\
&=&\frac{F_{0}}{\sigma \zeta }\sum\limits_{i=2}^{N}Q_{mi}\frac{1-\exp
(-\sigma \lambda _{i}t)}{\lambda _{i}}Q_{im}^{-1}.  \nonumber
\end{eqnarray}

\section{Numerical Evaluation of the Stretching of comb-Rings}

In this Section we evaluate the stretching $\left\langle \delta
Y_{m}(t)\right\rangle $ of comb-rings structures under external fields. For
this we start from comb molecules, consisting of a backbone made up of $%
M_{1} $ monomers, to which we attach linear chains of $M_{2}-1$\ monomers
each. For simplicity, we apply periodic boundary condition {\em to the
backbone}, or, equivalently, we close the backbone into a ring. The
situation is depicted in Fig. 1 where we have $M_{1}=9$ and $M_{2}=4$. As a
further simplification we set \ $M_{1}=M_{2}=M$ and choose $M$ to be $50$,
by which we have $N=$ $2500$. Then we diagonalise the corresponding ${\bf A}$%
{\bf \ }matrices using the program MATLAB and determine their eigenvalues
and eigenvectors. In terms of the GGS-model considered here, these
quantities are sufficient to allow to determine both the position dependent $%
\left\langle \delta Y_{m}(t)\right\rangle $, see Eq. (\ref{L}), as well as
the average of Eq. (12) with respect to $m$: 
\begin{equation}
\left\langle \left\langle \delta Y(t)\right\rangle \right\rangle =\frac{1}{N}%
\sum\limits_{m=1}^{N}\left\langle \delta Y_{m}(t)\right\rangle =\frac{F_{0}}{%
\sigma N\zeta }\sum\limits_{i=2}^{N}\frac{1-\exp (-\sigma \lambda _{i}t)}{%
\lambda _{i}},  \label{M}
\end{equation}
see Eq. (\ref{G}).

As discussed in previous works, the intermediate time regime of Eq. (\ref{L}%
) and (\ref{M}), namely $1/\sigma \lambda _{\max }\leq t\leq 1/\sigma
\lambda _{\min }$, (where $\lambda _{\min }$ and $\lambda _{\max }$ denote
the minimal nonvanishing and the maximal eigenvalue of the set $\left\{
\lambda _{i}\right\} )$ is particularly revealing of the underlying
GGS-topology\cite{BKB00,R53,FSB99,SFB99, KBB01, JSB00, GB01}. The same
holds, of course, for $\left\langle Y_{m}(t)\right\rangle $ and $%
\left\langle \left\langle Y(t)\right\rangle \right\rangle $, see Eq (\ref{Fa}%
) and Eq. (\ref{G}). Thus linear chains display in the time-interval
considered simple scaling with time, 
\begin{equation}
\left\langle \left\langle Y(t)\right\rangle \right\rangle \sim t^{\gamma }
\label{N}
\end{equation}
with $\gamma =1/2$ for Rouse-type GGS\cite{BKB00, DE86, KBB01}; this
behaviour is associated to anomalous diffusion and may be modelled through
fractional differential expressions\cite{FSB99, SFB99, GK01}. On the other
hand, already the behaviour of star-molecules\cite{BKB00, KBB01} is more
complex than Eq. (\ref{N}); dendrimers, hyperbranched polymers and networks
display intricate $\left\langle \left\langle Y(t)\right\rangle \right\rangle 
$ and $\left\langle \left\langle \delta Y(t)\right\rangle \right\rangle $\
forms\cite{KBB01, JSB00, GB01}.

Our aim is now to focus on the dependence of the stretching $\left\langle
\delta Y_{m}(t)\right\rangle $, Eq. (\ref{L}) on $m.$ As is evident from the
inherent symmetry of the model this dependence reflects the distance of the
bead on which the force acts from the ring backbone. We will also compare $%
\left\langle \delta Y_{m}(t)\right\rangle $ to its average over $m$, namely
to $\left\langle \left\langle \delta Y(t)\right\rangle \right\rangle $, Eq. (%
\ref{M}). The reason is that in several instances (for linear chains, for
fractals and also in part for dendrimers and for small-world-network
structures) Eq. (\ref{M}) was found\cite{BKB00, FSB99, SFB99, JSB00} to be a
qualitatively rather good description of the overall process. As will appear
evident in the following, ring-combs show as a function of time a stretching
which directly reflects the distance from the backbone of the bead on which
the external force acts; in fact, this effect is so large that Eq. (\ref{M})
ceases to be a qualitatively good description of the overall process. This
fact has profound implications; it means that stretching is a much more
revealing experimental method than macroscopic mechanical manipulations,
which are related to Eqs. (\ref{J}) and (\ref{K}); we will return to this
point in the next Section, after we present our evidence.

We start by presenting in Fig. 2 $\left\langle \delta Y_{m}(t)\right\rangle $%
, Eq. (\ref{L}) for short times. For the plot we use dimensionless units, so
that we set $\sigma \equiv 1$ and $F_{0}/\zeta \equiv 1$. Because of
symmetry, we need only to focus on the distance of the ${\em m}$-th bead
from the backbone. Thus $m=1$ indicates that the bead considered belongs to
the backbone and $m=50$ means that the bead is at one of the tips of the
chain-segments which constitute the comb. We depict first the situation at
relatively short times, $0\leqslant t\leqslant 1,$ and display in Fig. 2 the
parametric dependence of $\left\langle \delta Y_{m}(t)\right\rangle $ on $m$
for $m=1,2,49$, and 50. Also given by a dashed line is the average $%
\left\langle \delta Y(t)\right\rangle $.

We note from the start that for fixed $t$ the quantity $\left\langle \delta
Y_{m}(t)\right\rangle $ increases monotonically with $m$. This is simply
understood on physical grounds, since acting farther away from the backbone
extends a larger part of the linear-chain segment which constitutes the
comb. Moreover, in the time interval considered in Fig. 2 there appear three
types of behavior, namely those connected with the backbone $(m=1)$, with
the tips $(m=50)$ and with the rest, given that the curves for points
internal to the comb region (here $m=2$ and $m=49$) coincide to a large
degree. Clearly, the distinction between the three types of beads is their
coordination number Z, which equals 3 for $m=1$, $2$ for $m\epsilon \left\{
2,3,...,49\right\} ,$ and 1 for $m=50$. The physics of this stage is also
clear: At short times each bead feels only its spring connections to its
nearest-neighbors; a larger number of neighboring beads renders the
extension under an equal force slower. Because of the large number, namely
48, of internal beads, the average $\left\langle \delta Y(t)\right\rangle $
follows closely the dynamics of the internal beads in the temporal range of
Fig. 2.

Turning now to the behavior at very long times, we plot in Fig. 3 the
situation for $M=50$ and $0\leqslant t\leqslant 10000$, where we display the
curves corresponding to $m=1,2,10,20,30,40,49$, and $50$. As becomes evident
by looking on the right side of the Figure, at long times the curves
saturate. Here the above-mentioned monotonicity of the dependence of $%
\left\langle \delta Y_{m}(t)\right\rangle $ on $m$ for fixed $t$ is clearly
evident. Evident, furthermore, is that the final extension of the object is
(to a very good approximation) proportional to the value of $m$, {\em i.e. }%
practically proportional to its distance from the backbone. This
proportionality, in fact, is getting more and more exact for increasing $m$.
Note that such changes in the extension of the comb would lead to drastic
changes in the electronic energy transfer between pairs of chromophores (one
of which should be attached near to the backbone and the other close to the
monomer which is pulled), given the strong dependence of the transfer rates
on the mutual distance between the chromophores.

To render the dependence of the stretching on $m$ more explicit, we plot in
Fig. 4 $\ \left\langle \Delta _{m}(t)\right\rangle \equiv \left\{
\left\langle \delta Y_{m}(t)\right\rangle -\left\langle \delta
Y_{1}(t)\right\rangle \right\} /\left( m-1\right) $ for $m=2,10,20,30,40,49$
and $50$. Here one can notice first the very good scaling of all the curves
for $t$ large, their spread getting to be less than 3\%. The physical
explanation of this finding relies on a quasistatic picture: At very long
times the comb-ring diffuses as a whole through the solvent, its beads
moving with practically the same velocity; at such long times the influence
of the external force has propagated through the whole comb-ring, being
counterbalanced at the level of each bead by the friction acting on it. Then
the main contributions to $\left\langle \delta Y_{m}(t)\right\rangle $ are
those stemming from the springs of the arms on which the force acts; in this
way the situation is quite similar to that encountered in the case of
star-polymers \cite{BKB00, KBB01}: After its onset the force stretches one
after another the springs of the arm on which it acts, a deformation which
propagates until a bifurcation (or multifurcation) is reached, here the
ring, in the case of a star-polymer the center. After reaching the
bifurcation (or multifurcation) the force gets distributed vectorially along
different paths and its influence on $\left\langle \delta
Y_{m}(t)\right\rangle $ diminishes; from the point of view of the beads
belonging to one arm, the rest of the polymer behaves as a very heavy,
not-too-deformable mass. Hence in the stationary state mainly the springs
between the backbone and the site on which the force acts contribute to $%
\left\langle \delta Y_{m}(t)\right\rangle .$ Their stretching values are
very close and they enter additively into $\left\langle \delta
Y_{m}(t)\right\rangle $.

Returning now to Fig. 3 we notice that the $\left\langle \delta
Y_{m}(t)\right\rangle $ show strong differences as a function of $m$; we
also note that the average $\left\langle \delta Y(t)\right\rangle $ is not a
qualitative measure of the ''typical'' $\left\langle \delta
Y_{m}(t)\right\rangle $ behavior anymore. From Fig. 3 for instance, we find
that $\left\langle \delta Y(t)\right\rangle $ is close to $\left\langle
\delta Y_{m}(t)\right\rangle $ only for $m$ very close to $m=25$. Thus
situation differs here from previous findings for linear chains, and for
dendrimers\cite{BKB00, KBB01}. The reasons for the differences found are
complex: thus for linear chains and non-disperse samples (fixed $N$) the
differences between the $\left\langle \delta Y_{m}(t)\right\rangle $ are
rather limited (much less than an order of magnitude), while in the case of
dendrimers the beads at the tips are extremely numerous and the behavior of $%
\left\langle \delta Y_{m}(t)\right\rangle $ due to them dominates the average%
$\left\langle \delta Y(t)\right\rangle $.

Already from Fig. 3 we can notice that the ''approach to equilibrium'', {\em %
i.e.} reaching the steady-state extension happens differently for different $%
m$. The situation is underscored by Fig. 4, where $\left\langle \Delta
_{m}(t)\right\rangle $ highlights this effect: Thus the curve $\left\langle
\Delta _{2}(t)\right\rangle $ raises quickest from 0 as $t$ increases, a
behavior fallowed then by $\left\langle \Delta _{10}(t)\right\rangle .$ We
see that while the curves for $m=2$ and $10$ reach 80\% of the steady-state $%
\left\langle \Delta _{m}(t)\right\rangle $ for times around $200$ and $1200$
the curves whose $m$ are larger than $10$ reach their corresponding 80\%
values at later times. We interpret this as being again due to the
underlying dynamical extension processes, which let the $\left\langle \delta
Y_{m}(t)\right\rangle $ depend qualitatively on $m$. To analyze the
intermediate region between short and long times more carefully, with the
intention to uncover possible transient scaling behavior we plot in Fig. 5 $%
\left\langle \delta Y_{m}(t)\right\rangle $ {\em vs. }$t$ in double
logarithmic scales, focussing on the $10\leqslant t\leqslant 1000$ regime.
Depicted are again the curves for $m=1,2,10,20,30,40,49,$ and $50$, as well
as the averaged curve $\left\langle \delta Y(t)\right\rangle $. We remark at
once that the average scales quite nicely, which allows to approximate it in
this range by $\left\langle \delta Y(t)\right\rangle =ct^{\gamma }$, with $%
c=0.57$ and $\gamma =0.50$. The $\left\langle \delta Y_{m}(t)\right\rangle $
curves show slight deviations from scaling; however, where we to approximate
their behavior in Fig. 5 by straight lines, we would assign them $\gamma $%
-parameters ranging from $g=0.24$ for $m=1$ to $\gamma =0.53$ for $m=50$.
Without attributing excessive importance to these $\gamma $-values, we
remark only that they allow to quantify the $m$-dependence of the
stretching, $\left\langle \delta Y_{m}(t)\right\rangle .$ Furthermore $%
\gamma =0.5$ is the value obtained by stretching a macromolecular chain in
the Rouse-domain\cite{SFB99}; deviations from this value are observable at
very short times, where the motion of a single bead is ballistic, and hence $%
\gamma =1$, and also at very short times, where we have saturation and hence 
$\gamma \simeq 0$. One can view the general situation as being mainly
determined by the behavior of the chain-like segments; on top of it the
curve for $m=1$ mirrors the onset of the crossover to saturation, whereas
that for $m=50$ is more on the side of the short and medium times.

\section{Conclusions}

In this article we have focussed on the stretching of ring-combs in external
fields and have displayed, using the generalization by the Rouse-model to
Gaussian structures \cite{SB95} the dynamical unfolding of ring-combs when
one of their beads is directly experiencing a pulling force. A physical
realization would involve either having the corresponding bead charged, and
the ring-comb being exposed to an external electrical field or, more in line
with modern micromanipulation techniques, acting on the bead through optical
tweezers or by magnetic means. Distinct from situations involving
linear-chains, fractals\cite{SFB99, Sch98, FSB99, BJ02}, or even dendrimers%
\cite{KBB01, BKB00, F97}, comb-like macromolecules display a high
sensitivity of their response to the distance from the backbone of the bead
on which the external force acts. This precludes the theoretical use of
quantities averaged over all beads in the description of the response; in
mathematical terms, comb-like macromolecules require, besides the knowledge
of the eigenvalues of the connectivity matric {\bf A}, also the knowledge of
the corresponding eigenvectors. In descriptions that introduce other
features (i.e. the hydrodynamic interactions, the excluded volume, the
stiffness of the chains) the mathematical aspects are, in general, more
complex; given our experience with dendrimers, \cite{BKB00, KBB01} however,
we expect all the above qualitative statements (based on the Rouse model) to
stay correct. One has to be well-aware of the fact that a quantitative
comparison to experiments must take into account additional features.

We hasten to note that this has profound implications, given that such
usually measured mechanical responses such as the storage modulus $G^{\prime
}(\omega )$ and the loss modulus $G^{\prime \prime }(\omega )$, Eqs. (\ref{J}%
) and (\ref{K}), also involve the eigenvalues, but not the eigenvectors of 
{\bf A}. This is in line with $G^{\prime }(\omega )$ and $G^{\prime \prime
}(\omega )$ being macroscopically (and hence averaged) observable
quantities. Comb-like structures, on the other hand, display new dynamical
features, when probed by micromanipulation techniques. Thus, as shown in the
previous Section, a quantity such as $\left\langle \delta
Y_{m}(t)\right\rangle $ discloses a whole series of topological features in
its temporal evolution: $\left\langle \delta Y_{m}(t)\right\rangle $ depends
at short times mainly on $Z,$ the functionality $Z$ (number of connected
beads) of site $m$; at very long times $\left\langle \delta
Y_{m}(t)\right\rangle $ reflects the distance of bead $m$ from the ring
backbone; furthermore, at intermediate times, $\left\langle \delta
Y_{m}(t)\right\rangle $ may depend algebraically (as a power-law) on time, $%
\left\langle \delta Y_{m}(t)\right\rangle \sim t^{\gamma }$, where again $%
\gamma $ is $m$-dependent. This intermediate regime may become even more
rich as the size and the cross linking of the studied macromolecules
increase. We hence view applying micromanipulation techniques on
comb-molecules as a very worthwhile means of experimental study and suggest
to combine them with fluorescence techniques; from the point of view of
theory it may be valuable to extend our investigations to other
macromolecular entities with complex topologies. 

\section{Acknowledgement}

The support of the Deutsche Forschungsgemeinschaft, of the Fonds der
Chemischen Industrie and of the GIF are gratefully acknowledged.

\begin{center}
Figure captions
\end{center}

{\bf Figure 1 }

Model of comb-ring with $M_{1}=9$ beads on the ring and with attached chains
of $M_{2}-1=3$ beads each.

{\bf Figure 2}

Plot of $\left\langle \delta Y_{m}(t)\right\rangle $, Eq. (12), as a
function of time for several $m$ values, $m=1,2,49,$ and $50$ from below.
Given through a dashed line is also the average $\left\langle \left\langle
\delta Y(t)\right\rangle \right\rangle $, Eq. (13). The axes are in
dimensionless units and we set $\sigma =1$ and $F_{0}/\zeta =1$, see text
for details.

{\bf Figure 3}

Same as Fig. 2, for $m=1,2,10,20,30,40,49,$ and $50$ from below.

{\bf Figure 4}

Plot of $\left\langle \Delta _{m}(t)\right\rangle \equiv \left\{
\left\langle \delta Y_{m}(t)\right\rangle -\left\langle \delta
Y_{1}(t)\right\rangle \right\} /(m-1)$ as a function of $t$, in order to
display scaling at long times. The values of $m$ are 2,10,20,30,40,49, and
50. The two curves on the left side of the Figure belong to $m=2$ and to $%
m=10$, in this order, from above.

{\bf Figure 5}

Plot of $\left\langle \delta Y_{m}(t)\right\rangle $ vs. $t$ in double
logarithmic scales; the $m$ values are as in Fig. 3.


\newpage

\begin{references}
\bibitem{T75}  L.R.G. Treloar, {\em The Physics of Rubber Elasticity,}
Clarendon, Oxford, 1975.

\bibitem{ME92}  J.E. Mark and B. Erman eds., {\em Elastomeric Polymer
Networks, }Prentice Hall, Englewood Cliffs, NJ, 1992.

\bibitem{LSB99}  S. Lay, J.-U. Sommer, and A. Blumen, J. Chem. Phys. 110
(1999) 12173.

\bibitem{BKB00}  P. Biswas, R. Kant, and A. Blumen, Macromol. Theory Simul 9
(2000) 56.

\bibitem{PSLC}  T.T. Perkins, D.E. Smith, R.G. Larson, and S. Chu, Science
268 (1995) 83.

\bibitem{W95}  D. Wirtz, Phys. Rev. Lett. 75 (1995) 2436.

\bibitem{QBC97}  S.R. Quake, H. Babcock, and S. Chu, Nature (London) 388
(1997) 151.

\bibitem{GK01}  R. Granek and J. Klafter, Europhys. Lett. 56 (2001) 15.

\bibitem{KBB01}  R. Kant, P. Biswas, and A. Blumen, J. Chem. Phys. 114
(2001) 2430.

\bibitem{Sch98}  H. Schiessel, Phys. Rev. E 57 (1998) 5775.

\bibitem{FSB99}  C. Friedrich, H. Schiessel, and A. Blumen in: {\em Advances
in the Flow and Rheology of Non-Newtonian Fluids}, D. A. Siginer, D. DeKee,
R.P. Chhabra, Eds., Elsevier, Amsterdam 1999, pp. 429.

\bibitem{SFB99}  H. Schiessel, C. Friedrich, and A. Blumen in: {\em %
Applications of Fractional Calculus in Physics}, R. Hilfer, Ed., World
Scientific, Singapore 2000.

\bibitem{BJ02}  A. Blumen and A. Jurjiu, J. Chem. Phys. 116 (2002) 2636.

\bibitem{ME83}  J.E. Martin and B. E. Eichinger, Macromolec. 16 (1983) 1345.

\bibitem{RT87}  J. Roovers and P. M. Toporowski, Macromolec. 20 (1987) 2300.

\bibitem{L91}  J.E.G. Lipson, Macromolec. 24 (1991) 1327.

\bibitem{YMcLS94}  T. A. Yurasova, T. C. B. McLeish and A. N. Semenov,
Macromolec. 27 (1994) 7205.

\bibitem{F99}  J.J. Freire, Adv. Polym. Sci. 143 (1999) 35.

\bibitem{CR92}  D. Cassi and S. Regina, Modern Phys. Lett. B6 (1992) 1397.

\bibitem{BCV99}  R. Burioni, D. Cassi and A. Vezzani, J.Phys. A 32 (1999)
5539.

\bibitem{BCV00}  R. Burioni, D. Cassi and A. Vezzani, Eur.Phys. J. B 15
(2000) 665.

\bibitem{KB84}  J. Klafter and A. Blumen, J. Chem. Phys. 80 (1984) 875.

\bibitem{RB89}  A.K. Roy and A. Blumen, J. Chem. Phys. 91 (1989) 4353.

\bibitem{BKK97}  A. Bar Haim, J. Klafter, and R. Kopelmann, J.\ Amer. Chem.
Soc. 119 (1997) 6197.

\bibitem{SMB98}  I.M. Sokolov, J. Mai, and A. Blumen, J. Lumin. 76-77 (1998)
377.

\bibitem{PSK98}  T. Palszegi, I.M. Sokolov, and H.F. Kauffmann, Macromolec.
31 (1998) 2521.

\bibitem{W99}  S.E. Webber, Macromolec. Symp. 143 (1999) 359.

\bibitem{Y00}  E.K.L. Yeow et al., J. Phys. Chem. B 104 (2000) 2596.

\bibitem{LP00}  L.M.S. Loura and M. Prieto, J. Phys. Chem. B 104 (2000) 6911.

\bibitem{RW01}  Y. Rharbi and M.A. Winnik, Macromolec. 34 (2001) 5238.

\bibitem{FSW01}  J.P.S. Farinha, J.G. Spiro and M.A. Winnik, J. Phys. Chem.
B 105 (2001) 4879.

\bibitem{BBSM01}  E.N. Bodunov, M.N. Berberan-Santos, J.M.G. Martinho, Chem.
Phys. 274 (2001) 243.

\bibitem{BBM}  E.N. Bodunov, M.N. Berberan-Santos, J.M.G. Martinho, Chem.
Phys. Lett. 340 (2001) 137.

\bibitem{R53}  P.E. Rouse, J. Chem. Phys. 21 (1953) 1272.

\bibitem{DE86}  M. Doi and S.F. Edwards, {\em The Theory of Polymer
Dynamics, }Clarendon, Oxford, 1986.

\bibitem{GK94}  A.Yu. Grosberg and A.R. Khokhlov, {\em Statistical Physics
of Macromolecules}, AIP Press, New York, 1994.

\bibitem{SB95}  J.-U. Sommer and A. Blumen, J. Phys. A 28 (1995) 6669.

\bibitem{EM78}  B.E. Eichinger and J.E. Martin, J. Chem. Phys. 69 (1978)
4594.

\bibitem{ME80}  J.E. Martin and B.E. Eichinger, Macromolecules 13 (1980) 626.

\bibitem{KMF90}  A. Kloczkowski, J.E. Mark, and H.L. Frisch, Macromolecules
23 (1990) 3481.

\bibitem{AG91}  G. Allegra and F. Ganazzoli, Prog. Polym. Sci. 16 (1991) 463.

\bibitem{GP92}  M. Guenza and A. Perico, Macromolecules 25 (1992) 5942.

\bibitem{F97}  R. La Ferla, J. Chem. Phys. 106 (1997) 688.

\bibitem{GB01}  A.A. Gurtovenko and A. Blumen, J. Chem. Phys. 115 (2001)
4924.

\bibitem{JSB00}  S. Jespersen, I.M. Sokolov, and A. Blumen, J. Chem. Phys.
113 (2000) 7652.

\newpage
\end{references}
\end{document}